\documentclass[useAMS,usenatbib]{mn2e}

\usepackage{graphicx,amsmath}
\newcommand{\pbj}{\ifmmode{P_{\rm B1}}\else$P_{\rm B1}$\fi}
\newcommand{\fbj}{\ifmmode{f_{\rm B1}}\else$f_{\rm B1}$\fi}

\title[Intriguing triple-mode RR~Lyrae star with period doubling]{Intriguing triple-mode RR~Lyrae star with period doubling}
\author[R. Smolec et al.]
{R. Smolec$^{1}$\thanks{E-mail: smolec@camk.edu.pl}, I. Soszy\'nski$^{2}$, A. Udalski$^{2}$, M.K. Szyma\'nski$^{2}$, P. Pietrukowicz$^{2}$\and
J. Skowron$^{2}$, S. Koz\l{}owski$^{2}$, R. Poleski$^{2,3}$, P. Moskalik$^{1}$, D. Skowron$^{2}$\and 
G. Pietrzy\'nski$^{2,4}$, \L{}. Wyrzykowski$^{2,5}$, K. Ulaczyk$^{2}$ \& P. Mr\'oz$^{2}$\\
$^{1}$ Nicolaus Copernicus Astronomical Centre, Polish Academy of Sciences, Bartycka 18, 00-716 Warszawa, Poland\\
$^{2}$ Warsaw University Observatory, Al. Ujazdowskie 4, 00-478 Warszawa, Poland\\
$^{3}$ Department of Astronomy, Ohio State University, 140 W. 18th Ave., Columbus, OH 43210, USA\\
$^{4}$ Universidad de Concepci\'on, Departamento de Astronomia, Casilla 160-C, Concepci\'on, Chile\\
$^{5}$ Institute of Astronomy, University of Cambridge, Madingley Road, Cambridge CB3 0HA, UK
}

\begin{document}

\date{Accepted . Received ; in original form }

\pagerange{\pageref{firstpage}--\pageref{lastpage}} \pubyear{2014}

\maketitle

\label{firstpage}

\begin{abstract}
We report the discovery of an intriguing triple-mode RR~Lyrae star found in the OGLE Galactic bulge collection, OGLE-BLG-RRLYR-24137. In the OGLE catalog the star was identified as RRd star -- double-mode pulsator, pulsating simultaneously in the fundamental and in the first overtone modes. We find that third mode is excited and firmly detect its period doubling. Period ratios are not far from that expected for triple-mode -- fundamental, first and third overtone -- pulsation. Unfortunately, we cannot reproduce period ratios of the three modes with a consistent set of pulsation models. Therefore the other interpretation, that additional mode is non-radial, is also likely.
\end{abstract}

\begin{keywords}
stars: horizontal branch -- stars: oscillations -- stars: variables: RR~Lyrae -- stars: individual: OGLE-BLG-RRLYR-24137
\end{keywords}

\section{Introduction}\label{sec:intro}
%%%%%%%%%%%%%%%%%%%%%%%%%%%%%%%%%%%%%%%

Majority of the RR~Lyrae stars are single-mode radial pulsators, pulsating either in the fundamental mode (F mode, RRab stars) or in the first overtone mode (1O mode, RRc stars). Less frequent are double-mode pulsators, pulsating simultaneously in the fundamental and in the first overtone modes (RRd stars). Other interesting forms of double-mode pulsation were discovered recently. 

Double-mode, radial, fundamental and second overtone pulsation was reported in several stars observed from space by {\it CoRoT} and {\it Kepler}, and in one star observed from the ground \citep[e.g.][]{poretti,benko10,jurcsik_mwlyr}, for a review see \cite{pam13}. These stars form a tight progression in the Petersen diagram with period ratios clustering around $\sim\!0.59$. Several of these stars show the long-term quasi-periodic modulation of the fundamental mode -- the Blazhko effect. First overtone is not detected in these stars. We note that similar form of pulsation with one `intermediate' mode missing was detected in 1O+3O Cepheids \citep{ogle_freaks}.

Other interesting form of pulsation was discovered in RRc and in RRd stars, but not in RRab stars. Therefore, excitation of first overtone seems crucial in this group. The period of additional mode is shorter than the first overtone period, period ratios cluster around $\sim\!0.61$. This period ratio cannot be explained with excitation of two radial modes and therefore it is commonly assumed that additional mode is non-radial \citep{pamsm14}. Majority of these stars were discovered in space photometry gathered by {\it MOST}, {\it CoRoT} and {\it Kepler} \citep[e.g.][]{aqleo,pamsm14,szabo_corot}, few stars were detected from the ground \citep[e.g.][]{om09}. Only recently \cite{netzel} increased the number of known stars of this type by a factor of 6, analysing the OGLE-III observations of RRc and RRd stars of the Galactic bulge. Interestingly, in majority of the stars observed from space period doubling of additional non-radial mode is detected \citep[manifested through sub-harmonic frequencies in the frequency spectrum,][and references therein]{pamsm14}.

RRd stars with additional $\sim\!0.61$ non-radial mode are triple-mode radial--non-radial pulsators. So far triple-mode radially pulsating RR~Lyrae star was not detected. We note that a few radial, triple-mode classical Cepheids are known, either of F+1O+2O type or of 1O+2O+3O type \citep{mkm04,ogle_freaks,ogle_cep_smc,poleski}.
 
Triple, or multi-mode pulsators are very important, as precisely measured pulsation periods strongly constrain the stellar model and allow asteroseismic investigation even in classical pulsators \citep{pamwd05}, provided pulsation modes are identified. Detection of period doubling effect is also very important as it provides more insight into pulsation properties of the stars. In many types of classical pulsators, period doubling is a dynamical phenomenon caused by the half-integer resonance between pulsation modes \citep{mb90}. The resonance provides further constraints on the models and allows better understanding of pulsation dynamics of the stars \citep[e.g.][]{kms11,blher}. We note that we do not understand the mechanism behind even the simplest form of multi-mode pulsation: double-mode radial pulsation. For a recent review see \cite{smolec14}.

In this paper we study the photometry of OGLE-BLG-RRLYR-24137, a triple-mode RR Lyrae star with period doubling effect. In the next Section we briefly describe the data available for the star and its analysis. Interpretation of the frequency spectrum is then provided in Section~\ref{sec:discussion} in which we also discuss the results. Short summary closes the paper.

\section{Observations and data analysis}\label{sec:methods}
%%%%%%%%%%%%%%%%%%%%%%%%%%%%%%%%%%%%%%%%%%%%%%%%%%%%%%%%%%%

\begin{figure}
\centering
\resizebox{\hsize}{!}{\includegraphics{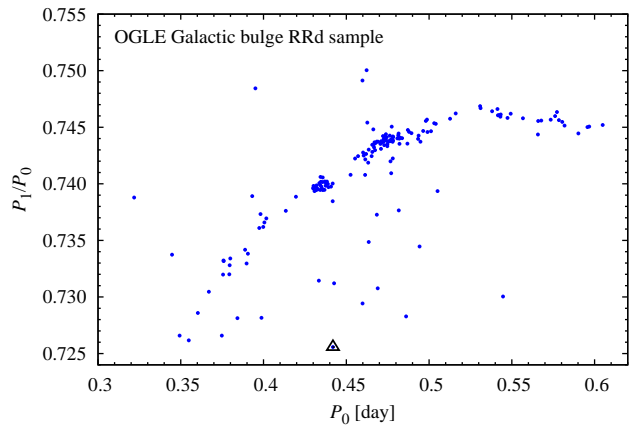}}
\caption{The Petersen diagram for RRd stars of the Galactic bulge OGLE collection. Investigated star is marked with triangle.}
\label{fig.pet}
\end{figure}

OGLE-BLG-RRLYR-24137 was discovered during the fourth, ongoing phase of the Optical Gravitational Lensing Experiment (OGLE), in the Galactic bulge field \citep{ogleiv_rrl_blg}. We refer the reader to \cite{ogleIII} for detailed description of the instrument setup and photometry reduction procedures. Altogether $481$ photometric epochs were collected in the $I$-band over five observing seasons. The number of data points per observing season vary from 56 to 130 and season length vary from $\sim$202\thinspace d to $\sim$256\thinspace d. Spectral window of the data is typical for the OGLE photometry of the Galactic bulge: strong daily aliases and one-year aliases are present. Fortunately, aliasing is not a source of confusion at any point of our analysis. Only $20$ observations were collected in the $V$-band and these data are not used in the analysis. We note that the mean brightness of the star and its color are typical for the Galactic bulge RR~Lyrae stars, but we admit that spread of these parameters in the bulge is very large, as compared e.g. to the Magellanic Cloud pulsators. The star was identified by \cite{ogleiv_rrl_blg} as RRd pulsator, although its period ratio is somewhat not typical, $P_1/P_0=0.7256$ -- lower than for majority of the RRd stars with a similar period of the fundamental mode -- see the Petersen diagram in Fig.~\ref{fig.pet}. We note that we have analysed in detail the stars deviating in the Petersen diagram in a separate publication \citep{smolec_rrd}. Majority of these stars show modulation of pulsation modes akin to the Blazhko effect. This in not the case for OGLE-BLG-RRLYR-24137, as we show in this paper.

\begin{figure*}
\centering
\resizebox{\hsize}{!}{\includegraphics{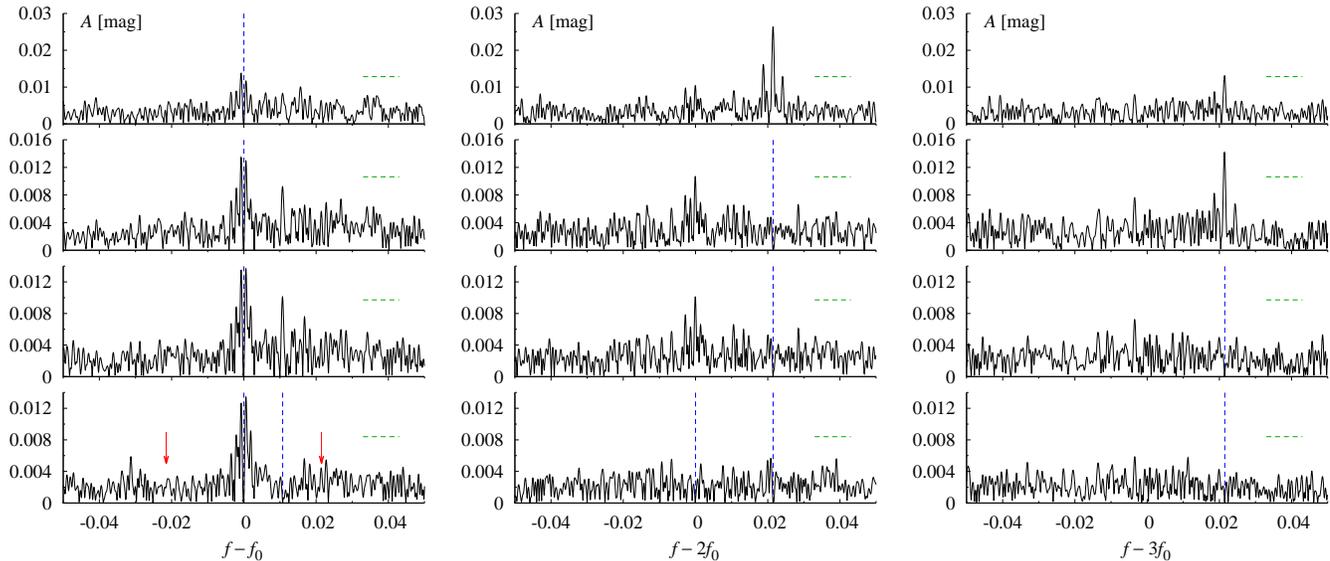}}
\caption{Prewhitening sequence for OGLE-BLG-RRLYR-24137. Left, middle and right panels show zoom-ins centered at $f_0$, $2f_0$ and $3f_0$, respectively. In the consecutive rows following frequencies were prewhitened: first row -- $f_0$ and $f_1$; second row --  as in the previous row plus $f_x$ and $f_0+f_1$; third row -- as in the previous row plus $f_0+f_x$ and $f_{\rm u}$; fourth row -- all detected frequencies from top part of Tab.~\ref{tab:freqs} were prewhitened. Prewhitened frequencies are marked with vertical dashed lines. Short horizontal line segments in the right part of each panel indicate the ${\rm S/N}=4$ level. Two red arrows in the bottom-left panel show the expected location of triplet components, in case pulsation is modulated with $P_{\rm B}=46.5$\thinspace d (see Discussion). }
\label{fig.fsp}
\end{figure*}

Data are analysed using standard successive prewhitening technique. Significant periodicities are identified with the help of discrete Fourier transform. At each step of the iterative procedure sine series with all identified frequencies is fitted to the data:
\begin{equation}
m(t)=m_0+\sum_{k=1}^N A_k\sin\big(2\pi f_k t + \phi_k\big)\,.\label{eq:ss}
\end{equation}
Amplitudes, phases and frequencies are adjusted by means of non-linear least square fit. Residuals from the fit are inspected for additional signals with the arbitrary signal-to-noise criterion, ${\rm S/N}\!>\!4$. We also accepted signals with slightly lower ${\rm S/N}$ values, provided they were located exactly at the linear frequency combination of the previously identified independent frequencies. Eleven frequencies identified this way are collected in the top part of Tab.~\ref{tab:freqs}, together with their amplitudes and two possible interpretations to be discussed in a moment. First two frequencies, $f_0$ and $f_1$, correspond to fundamental mode and first overtone as identified already in the OGLE catalog. Successive rows of Fig.~\ref{fig.fsp} illustrate the prewhitening process. 
\begin{table}
\centering
\caption{Significant frequencies detected in OGLE-BLG-RRLYR-24137 and their possible interpretations: triple-mode pulsation with period doubling (third column) or double-mode pulsation with modulation (fourth column). In the last row of the Table frequency detected \mbox{after} application of time-dependent prewhitening is given.}
\label{tab:freqs}
\begin{tabular}{lrrr}
frequency & amplitude & triple-mode & modulation \\
\hline
2.262844 &  0.0800 &         $f_0$  & $f_0$ \\
3.118717 &  0.0607 &         $f_1$  & $f_1$ \\
4.547226 &  0.0293 &         $f_x$  & $2f_0+f_{\rm B}$ \\
6.810070 &  0.0150 &     $f_0+f_x$  & $3f_0+f_{\rm B}$ \\
5.381561 &  0.0150 &     $f_0+f_1$  & $f_0+f_1$ \\
0.834335 &  0.0116 & $f_0+f_1-f_x$  & $f_1-f_0-f_{\rm B}$ \\
2.273613 &  0.0104 &     $0.5f_x$   & $f_0+0.5f_{\rm B}$ \\
4.525689 &  0.0100 &       $2f_0$   &  $2f_0$  \\
3.709679 &  0.0128 &    $f_{\rm u}$  &  $f_{\rm u}$ \\
9.928788 &  0.0077 & $f_0+f_1+f_x$  & $3f_0+f_1+f_{\rm B}$   \\
3.097180 &  0.0072 & $2f_0+f_1-f_x$ & $f_1-f_{\rm B}$    \\
\hline
6.237434 &  0.0078 &        $2f_1$ & $2f_1$ \\
\hline
\end{tabular}
\end{table}

After prewhitening with all frequencies listed in the top part of Tab.~\ref{tab:freqs} (4 independent frequencies, 7 linear combinations), a significant signal remains in the frequency spectrum at the frequency of the fundamental mode. It is unresolved with $f_0$ (frequency separation is smaller than $2/T$, $T$ -- data length). Such signal is a signature of non-stationary nature of the fundamental mode, a likely slow (not resolved within available data length) variation of its amplitude and/or phase. It increases the noise level in the Fourier transform and may hide additional peaks. To get rid of this signal and search for additional frequencies we conducted the time-dependent prewhitening on a season-to-season basis. This technique is described by \cite{pamsm14} and is based on the time-dependent Fourier analysis \citep{kbd87} followed by the data prewhitening. In a nutshell, data were divided into five subsets, each corresponding to one observing season. Next, we fitted the sine series to each of the subsets independently, with all frequencies listed in the top part of Tab.~\ref{tab:freqs}, adjusting the amplitudes and phases, but keeping the frequencies fixed (as determined from the just described analysis of the full data). Any possible frequency changes are then reflected in the variation of respective phases. We stress that number of data points and season lengths are sufficient to conduct such analysis (see first paragraph of this Section).

\begin{figure}
\centering
\resizebox{\hsize}{!}{\includegraphics{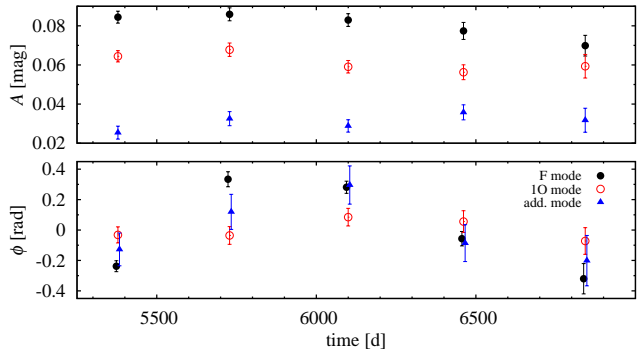}}
\caption{Season-to-season variation of amplitudes (top panel) and phases (bottom panel) of the fundamental mode (filled circles), first overtone mode (open circles), and of the frequency denoted in Tab.~\ref{tab:freqs} as $f_x$ (triangles). Phases are shifted vertically so their mean values are 0. Also for clarity, in each season the three phases where shifted horizontally by $\pm 5$\thinspace d, as otherwise the error-bars overlap.}
\label{fig.tdfd}
\end{figure}

Results of the subset analysis are presented in Fig.~\ref{fig.tdfd}, in which upper panel shows the amplitude variation of the fundamental and first overtone modes, as well as of the frequency denoted in Tab.~\ref{tab:freqs} by $f_x$, interpreted in the following sections as additional pulsation mode. The lower panel shows the phase variations. The non-stationarity of the fundamental mode is clearly visible: its phase strongly vary and also its amplitude change. The variation of other two modes is also clear but less pronounced.

 Residuals from the five independent fits are merged and analysed. Non-stationary signals are removed, provided that associated variation is not significant within one observing season. This is indeed the case. The noise level in the Fourier transform drops, which is also due to the levelling of possible season-to-season zero-point differences in the photometry. In the data we find one additional significant signal listed in the bottom part of Tab.~\ref{tab:freqs} (last row).

\section{Discussion}\label{sec:discussion}
%%%%%%%%%%%%%%%%%%%%%%%%%%%%%%%%%%%%%%%%%%%%

\subsection{Additional mode or modulation?}
%%%%%%%%%%%%%%%%%%%%%%%%%%%%%%%%%%%%%%%%%%%

After prewhitening the data with frequencies corresponding to the fundamental and first overtone modes, significant signal appears in the frequency spectrum close to $2f_0$ (top row in Fig.~\ref{fig.fsp}). Two interpretations are possible. Either it is an independent pulsation mode, which we denote by $f_x$, or it is a signature of modulation of the fundamental mode. In the latter case close multiplet structures should be detected at the frequency of the fundamental mode and its harmonics, $kf_0$. In the ground-based data these are typically triplets, which may be strongly asymmetric and appear as doublets \citep[see e.g.][]{alcock_rrab,benko11}. The frequency separation between the multiplet/doublet components corresponds to modulation frequency and its inverse corresponds to modulation period. Here we denote the modulation frequency as $f_{\rm B}$. Hence our two interpretations for the discussed frequency are $f=f_x$ (independent pulsation mode) or $f=2f_0+f_{\rm B}$ (modulation doublet; resulting modulation period is $46.5$\thinspace d). Except one ($f_{\rm u}$), all other frequencies detected in the frequency spectrum may be expressed as linear combinations of $f_1$, $f_2$ and $f_x$, or as linear combinations of $f_1$, $f_2$ and $f_{\rm B}$. These two possible interpretations are collected in the third and fourth columns of Tab.~\ref{tab:freqs}. The only frequency that cannot be represented this way is denoted as $f_{\rm u}$. The period ratios, $P_{\rm u}/P_0\approx 0.61$ and $P_{\rm u}/P_1\approx 0.84$ are far from those expected for radial modes. The first period ratio is encountered in RR Lyrae stars, but in relation to first overtone period, not the fundamental mode period (see Section~\ref{ssec:addmode} for more details). We do not find any frequency combinations involving this frequency, consequently we do not have a proof that it originates from OGLE-BLG-RRLYR-24137. Detection of $f_{\rm u}$ is certain (${\rm S/N}=5.0$). This signal may correspond to unresolved blend or e.g. other, non-radial pulsation mode. We ignore this signal in the following discussion.

Which of the two interpretations is correct? We argue that the modulation scenario is unlikely. Our arguments are:
\begin{itemize}
\item In the case of modulation we expect equidistant triplets, which are not detected. This is not a strong point however, as triplets are often highly asymmetric and may appear as doublets.
\item The highest signals corresponding to modulation are detected at $2f_0+f_{\rm B}$ and $3f_0+f_{\rm B}$ but we do not find any significant signal at $f_0+f_{\rm B}$ or $f_0-f_{\rm B}$ (see bottom row in Fig.~\ref{fig.fsp}, arrows mark the expected location of the modulation side peaks)! The side peaks at harmonic frequencies may be slightly higher than at $f_0$ \citep{benko11} but here signal at $f_0\pm f_{\rm B}$ is missing. In addition side peaks are much higher than harmonics itself; even more, the harmonic at $3f_0$ is not detected at all (bottom right panel in Fig.~\ref{fig.fsp}).
\item A close and significant signal at $f_0$ is detected, but at a separation corresponding to $f_{\rm B}/2$ (third row in Fig.~\ref{fig.fsp}). If we assume that this signal corresponds to a true modulation frequency, i.e. $f_{\rm B}'=f_{\rm B}/2$ we have even more severe problem with modulation scenario as incomplete quintuplets appear in the spectrum then.
\item Incomplete quintuplets are detected at $2f_0$ and $3f_0$. These are extreme quintuplets with all low frequency components missing (only $2f_0+2f_{\rm B}'$ and $3f_0+2f_{\rm B}'$ are detected).
\item Detection of modulation peak at higher order radial mode combination frequency, at $3f_0+f_1$, is also suspicious, taking into account that $3f_0$ itself is not detected. 
\end{itemize}    

Based on these arguments we conclude that modulation scenario is unlikely. The second scenario, excitation of additional, independent pulsation mode is most likely. Interpretation of frequency spectrum faces no difficulty in this case. Detected signals are low-order linear combinations of the three modes. This is the interpretation we adopt in the remaining of this paper.

\subsection{Period doubling of the additional mode.}
%%%%%%%%%%%%%%%%%%%%%%%%%%%%%%%%%%%%%%%%%%%%%%%%%%%%

In the frequency spectrum we firmly detect a sub-harmonic of $f_x$, a significant signal at $f=0.5f_x$. Deviation from exact sub-harmonic is negligible, $f-0.5f_x=-0.00007$\thinspace c/d (estimated by the non-linear least-square fit of the time-series). We do not find other sub-harmonic frequencies. The presence of sub-harmonic frequency is a signature of period doubling of $f_x$ \citep[e.g.][]{blher}. Indeed, we can directly see the effect in the disentangled light curve corresponding to $f_x$. 

\begin{figure}
\centering
\resizebox{\hsize}{!}{\includegraphics{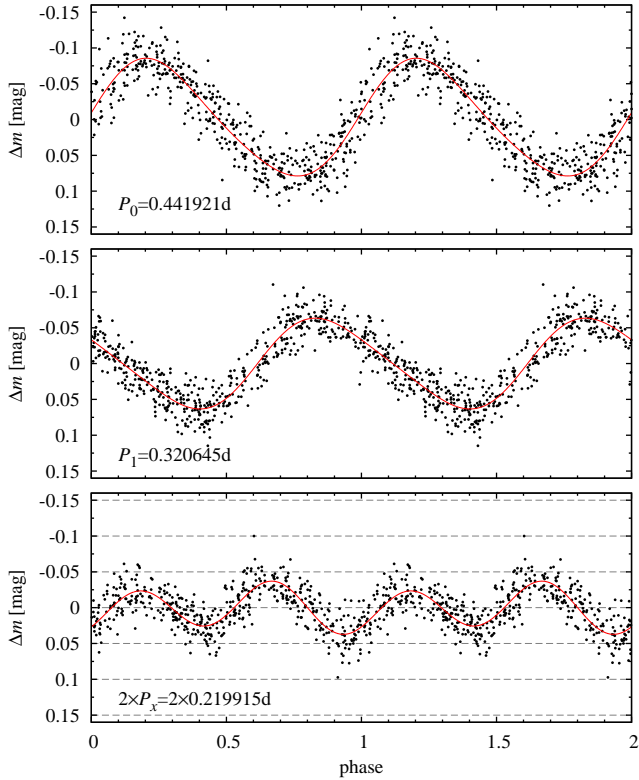}}
\caption{Disentangled light curves for different pulsation modes in OGLE-BLG-RRLYR-24137: top panel -- fundamental mode, middle panel -- first overtone, bottom panel -- additional mode, suspected third overtone with period doubling effect.}
\label{fig.curves}
\end{figure}

The phased light curves corresponding to the three pulsation modes are plotted in Fig.~\ref{fig.curves}. To get the light curve of the fundamental mode we first subtracted from the data a sine series with all frequencies listed in Tab.~\ref{tab:freqs} except those corresponding to $f_0$ and $2f_0$, and next phased the data with $P_0=1/f_0$. The light curves for the first overtone and for the additional mode were extracted in a slightly different manner. In order to fully remove the non-stationary fundamental mode from the data, we used time-dependent prewhitening on a season-to-season basis, as described above. This time however, we fitted the sine series to each of the subsets without $f_1$ and $2f_1$ terms (to get the first overtone light curve) or without $f_x$ and $0.5f_x$ terms (to get the light curve of the additional mode). The resulting prewhitened data were phased either with $P_1=1/f_1$ or with $2P_x=2/f_x$.

\begin{figure}
\centering
\resizebox{\hsize}{!}{\includegraphics{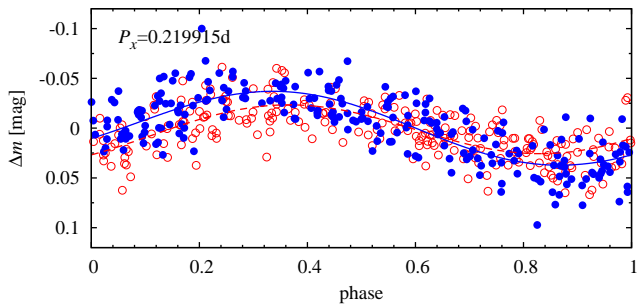}}
\caption{Disentangled light curve of the suspected third overtone, folded with $P_x$. Points corresponding to even and odd pulsation cycles are plotted with different symbols/colors.}
\label{fig.eo}
\end{figure}

Period doubling effect is weak, but visible in the light curve of the additional mode. We see alternating deep and shallow minima. The differences are also visible at maximum light. Fourier fit (over-plotted) confirms the effect clearly. In Fig.~\ref{fig.eo} we present a different visualization of the effect: even and odd pulsation cycles are plotted with different symbols. The effect is well visible, but certainly not very strong.

\begin{figure}
\centering
\resizebox{\hsize}{!}{\includegraphics{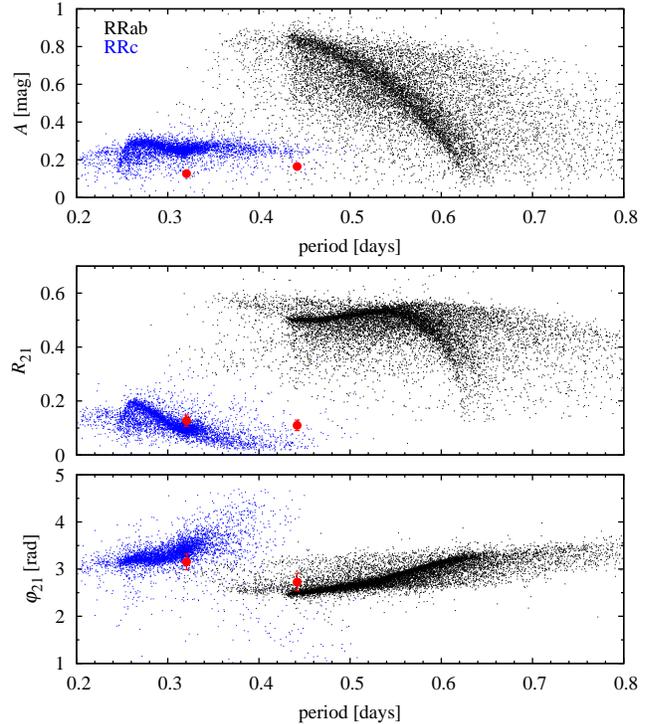}}
\caption{Peak-to-peak amplitude (top) and lowest order Fourier decomposition parameters, $R_{21}$ and $\varphi_{21}$ (middle and bottom), for disentangled fundamental and first overtone light curves, compared with parameters of RRab and RRc stars from the OGLE-III Catalog of Variable Stars \citep{ogle_rrl_blg}.}
\label{fig.fps}
\end{figure}

As mentioned in the Introduction, our star is not a typical RRd star because of low $P_1/P_0$ period ratio (Fig.~\ref{fig.pet}). It is hence interesting to investigate whether light curves of the fundamental and first overtone modes are typical for RR~Lyrae stars. To this aim, we compare their lowest order Fourier decomposition parameters, $R_{21}=A_2/A_1$ and $\varphi_{21}=\phi_2-2\phi_1$, with the parameters for RRab and RRc stars of the OGLE-III catalog -- Fig.~\ref{fig.fps}. Amplitudes of both modes are lower than for majority of the RRab/RRc stars, which is expected (two modes saturate the driving mechanism). Amplitude of the first overtone is only slightly lower, while amplitude of the fundamental mode is significantly lower. Consequently its $R_{21}$ value is also very low. On the other hand, the Fourier phase for the fundamental mode is typical for the fundamental mode pulsators. In the case of the first overtone mode, both amplitude ratio and Fourier phase are typical.

We also note that light curve displayed in the bottom panel of Fig.~\ref{fig.curves} resembles that observed for close (contact) binary systems. Such interpretation is unlikely for the discussed star, however. The orbital period would be slightly below $11$\thinspace hours then. A simple estimate based on the third Kepler law shows that it is not possible to fit an RR~Lyrae-type star inside a resulting tight orbit. Assuming total mass of the system in a range $0.25-5.0$ solar masses the resulting semi-major axes of the orbit would be in between $1.5$ and $4.2$ solar radii. The typical radii expected for RR~Lyrae-type stars are $4-6\,{\rm R}_\odot$. Similar radii are expected for the less massive relatives of RR~Lyrae stars that evolved in a tight binary system -- the binary evolution pulsators \citep[BEPs,][]{bep_nat,bep} \citep[$R=4.24\pm 0.24\,{\rm R}_\odot$ for the star discovered by][]{bep_nat}. We also note that additional variability depicted in the lower panel of Fig.~\ref{fig.curves} cannot result from a blend of RRd star and of a tight binary system, as we clearly detect four combination frequencies involving $f_x$, $f_0$ and $f_1$ (Tab.~\ref{tab:freqs}). Also, rotation effects are rather unlikely to be the source of the variability, as RR Lyrae stars are slow rotators \citep{preston}, while variability present in the lower panel of Fig.~\ref{fig.curves} occurs on a time-scale of less than a day.

\subsection{Nature of the additional mode and cause of period doubling.}\label{ssec:addmode}
%%%%%%%%%%%%%%%%%%%%%%%%%%%%%%%%%%%%%%%%%%%%%%%%%%%%%%%%%%%%%%%%%%%%%%%%

The period ratio between the additional mode and the first overtone, $P_x/P_1=0.686$, indicates that the additional mode may correspond to the radial third overtone, although the period ratio seem to high. To check it, we have computed a set of RR~Lyrae models with different masses, luminosities and metallicities in a large parameter range ($0.5<M/{\rm M}_\odot<0.75$, $30<L/{\rm L}_\odot<70$, $0.00004<Z<0.02$) covering the full instability strip. The models were computed with the Warsaw pulsation codes \citep{sm08} adopting OPAL opacity tables \citep{opal} and \cite{a04} solar abundance mixture. All models adopt $X=0.76$ (results do not depend strongly on the choice of $X$, on opacity tables or on solar mixture). These are envelope models with homogeneous chemical composition.

\begin{figure}
\centering
\resizebox{\hsize}{!}{\includegraphics{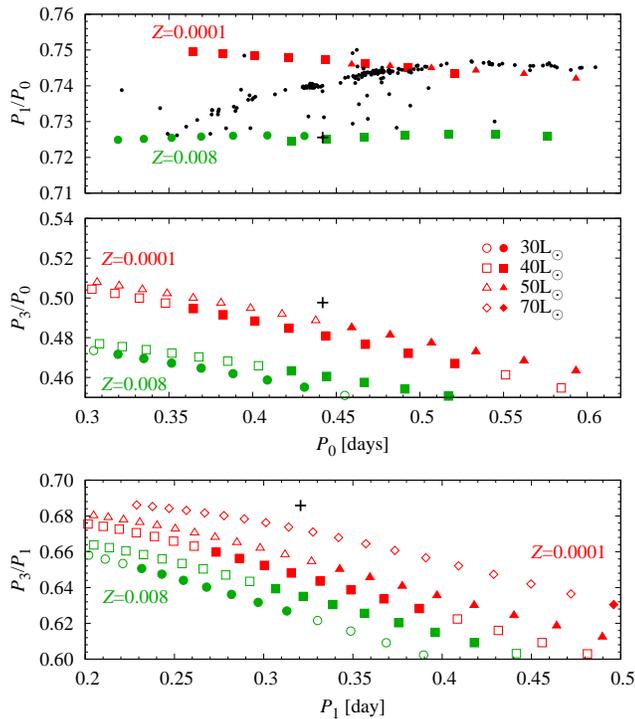}}
\caption{Period ratios for OGLE-BLG-RRLYR-24137 (cross) confronted with the pulsation models assuming that the additional mode is the third radial overtone (different symbols depending on model's luminosity, as indicated in the key). Filled symbols correspond to models in which both fundamental and first overtone modes are linearly unstable. All models have $M=0.6{\rm M}_\odot$.}
\label{fig.models}
\end{figure}

In the Petersen diagrams in Fig.~\ref{fig.models} we show only the models with period ratios closest to the period ratios determined for OGLE-BLG-RRLYR-24137. The three panels show, from top to bottom, the $P_1/P_0$, $P_3/P_0$ and $P_3/P_1$ radial mode period ratios. All these models have $M=0.6{\rm M}_\odot$, which is typical for RR~Lyrae stars and adopt either high metallicity ($Z=0.008$, green symbols) or low metallicity ($Z=0.0001$, red symbols), and have different luminosities as indicated in the key. Model sequences run horizontally across the instability strip with a $100$\thinspace K step in effective temperature. Filled symbols correspond to models in which fundamental and first overtone modes are simultaneously excited (their linear growth rates, $\gamma_0$ and $\gamma_1$, are positive), which is a necessary condition for (non-resonant) double-mode pulsation. Third overtone is stable in all the models (see the discussion in next paragraphs).

First, we note that not typical first overtone to fundamental mode period ratio may be explained by large metallicity of the star and relatively low luminosity \citep[which is expected at higher metallicities, see also][]{ogle_switch}. We note that Galactic bulge is characterized by large metallicity spread and large, solar-like metallicities are not rare \citep[e.g.][]{zoccali,bensby,ogle_rrl_blg}. However, with large metallicity we cannot reproduce the other period ratios, $P_x/P_0$ and $P_x/P_1$, they are much higher than high metallicity models predict. In order to approach these two period ratios we need very low metallicity values (which are more typical for RR~Lyrae stars). The observed $P_x/P_0$ ratio is not far from the $Z=0.0001$ $P_3/P_0$ model ratios. We face the most severe difficulty with $P_x/P_1$ period ratio which cannot be reproduced with the models. Reducing the metallicity further does not help. The closest models are characterised with low $M/L$ ratio, but are located far on the blue side of the instability strip.

We have also checked whether with low mass models, with masses in between $0.2-0.4$ solar masses, we can match the observed period ratios better. Such masses are expected for the binary evolution pulsators, for which light curves may resemble those of RR Lyrae stars \citep{bep_nat,bep}. Although we can match $P_0/P_1$ with high metallicity models, in the case of other period ratios, $P_x/P_0$ and $P_x/P_1$, the disagreement is even worse than in the case of RR~Lyrae models.

The radial mode interpretation faces the following difficulties:
\begin{itemize}
\item We cannot reproduce the three period ratios assuming consistent model parameters. $P_1/P_0$ requires high metallicity, other two period ratios require very low metallicity values
\item $P_x/P_1$ cannot be reproduced with the models. It is too high as compared to the model predictions for the third-to-first overtone period ratio.
\item Third overtone is always linearly stable in the models.
\end{itemize}
The last difficulty is actually a problem of excitation mechanism for the postulated third overtone, which cannot be non-resonant in such case, but may occur through a mode resonance. We note that even in the non-resonant scenario, the linear instability of two modes is only a necessary condition for the double-mode pulsation. Majority of the RRd stars form a well defined and tight sequence in the Petersen diagram (Fig.~\ref{fig.pet}), while models predict simultaneous instability of the two modes over a much larger area. A mode selection mechanism is in action here, which we do not understand, however \citep{smolec14}. We note that our star does not fit the progression formed by the majority of RRd stars. Why these and other stars deviate from the main progression is not known. In \cite{smolec_rrd} we suggest that these stars might be in a transient state following the RRab$\rightarrow$RRd mode switch. We note that similar to stars analysed in that paper, in OGLE-BLG-RRLYR-24137 the fundamental mode varies on a long time-scale, which is manifested by non-coherent signal in the frequency spectrum (Fig.~\ref{fig.fsp}) and well visible in Fig.~\ref{fig.tdfd}. Other explanation that comes to mind is a different internal structure of these stars e.g. the presence of discontinuities in the chemical profile along the radius which may strongly affect the computed period ratios of the radial modes. Without an analysis of the full evolutionary models this is speculative however.

The other explanation is that additional mode is non-radial. As described in the Introduction there is a new group of radial--non-radial RR~Lyrae pulsators with period ratios between the additional mode and the first overtone clustering around $\sim\!0.61$ (in between radial third-to-first overtone and fourth-to-first overtone period ratios). Some of these stars are genuine RRd pulsators \citep[][and references therein]{aqleo,chadid,pamsm14}. In majority of the stars observed from space, sub-harmonic frequencies, a signature of period doubling, are also detected \citep{pamsm14,szabo_corot}. These frequencies however, are often located slightly off the exact value of sub-harmonic frequency. There is no firm detection of sub-harmonic frequencies in stars detected in the ground-based observations. Marginal detection was reported in four out of 147 stars analysed by \cite{netzel} (OGLE-III data). Amplitudes of non-radial mode in the discussed group are typically few percent of the radial mode amplitude. To the contrary, in our star amplitude of additional mode is comparable to radial mode amplitudes. Sub-harmonic is firmly detected and located exactly where it should. We note that non-radial modes are linearly unstable in a large frequency range in the models \citep{hdk98,dc99}. The mode selection mechanism however, remains obscure (\cite{smolec14}, see also \cite{wd12}).

We conclude that the additional frequency corresponds to independent pulsation mode, but whether it is a radial mode or non-radial mode cannot be answered at the moment.

A firmly detected period doubling is most likely a resonant effect as first analysed by \cite{mb90}. It is hard to point which resonance may be responsible for the effect in the case of our star, as we cannot firmly identify the mode itself. We note that it is yet another discovery of period doubling in classical pulsators reported in the recent years, after discovery of period doubling in Blazhko RRab stars \citep{szabo10}, in BL~Her type star \citep{blher} and in the just discussed radial--non-radial pulsators with $\sim\!0.61$ period ratio \citep{pamsm14}.

\section{Summary}
%%%%%%%%%%%%%%%%%%%%%%%%%%%

OGLE-BLG-RRLYR-24137 is a triple-mode RR~Lyrae star. The two dominant modes correspond to the radial fundamental and first overtone modes. The corresponding period ratio is smaller than for majority of RRd stars with similar period, the star is not an exception however (Fig.~\ref{fig.pet}). The shape and Fourier parameters of the disentangled light curves support the RRd identification (Figs.~\ref{fig.curves} and \ref{fig.fps}). The third pulsation mode has comparable amplitude to the two radial modes. Most interestingly it undergoes a period doubling (Fig.~\ref{fig.curves}). Based on the data we have and model computations done, we cannot decide whether it is the radial third overtone mode or a non-radial mode. In the former case we cannot reproduce the observed period ratios with standard homogeneous envelope pulsation models. 

No doubt, this intriguing triple-mode star deserves more observation and study. It demonstrates the power of massive sky surveys in search for unique and interesting objects. We cannot judge whether it is an isolated case or first member of a new class of RR~Lyrae pulsators, but analysis of expected data for $\sim 10^5$ RR~Lyrae pulsators from OGLE-IV, of which photometry for more than $38\,000$ Galactic bulge stars was just published \citep{ogleiv_rrl_blg}, will help to resolve this issue.

\section*{Acknowledgments}
We are grateful to the anonymous referee for several valuable comments. This research is supported by the Polish National Science Centre through grant DEC-2012/05/B/ST9/03932 and by the Polish Ministry of Science and Higher Education through the program ``Ideas Plus'' award No. IdP2012 000162. The OGLE project has received funding from the European Research Council under the European Community's Seventh Framework Programme (FP7/2007-2013)/ERC grant agreement no. 246678 to AU.

\bsp

\label{lastpage}

\end{document}